**Predicting patient outcomes (TNBC) based on positions of cancer islands and CD8+ T cells using machine learning approach**
Guangyuan Yu, Xuefei Li, Herbert Levine

## Introduction

Machine learning method is being applied in cancer research. Heidari et al. used machine learning approach to predict the short-term breast cancer risk [1]. Agarap et al. compared six machine learning (ML) algorithms on Wisconsin Diagnostic Dataset for a binary prediction problem of benign tumor or malignant tumor [2]. Saltz et al. used convolutional neural network to generate the TIL maps of TCGA samples to study the spatial structure of tumors and showed the percentage of high infiltration for several kinds of cancer [3]. On the other hand, high $CD8^+$ T cell counts (both overall and inside cancer-cell islands) is associated with better patient outcome [4]. However, a cut-off of the T-cell count has to be selected manually to separate groups of patients. In this work, we propose a method to classify the small patch of triple-negative breast cancer (TNBC) tumor and use the overall percentage of "good" patches as a marker to predict the prognosis, which is an automatic method of prognosis and could also be used for other cancers.

## Methods

<u>Data pre-processing</u>
- Cancer cell islands are generated by InForm platform.
- Centroid of CD8+ T cells are generated by a MATLAB algorithm developed in house, which combines the information of cytosol generated by InForm and the signal of CD8.
- A JPG image with a size of (~20k x 20k pixels) is generated for each patient (24 in total). The size of each pixel is around 1um x 1um. In each JPG image, cancer cell islands are labelled as white and each CD8+ T cells is marked by x pixels around its centroid.
- Each image is then resized by a factor of 7.8, and then divided into grids of sub-images. Each sub-image has a size of 64 x 64 pixels.
- If the sub-image has very few white pixels, less than a quarter of the total area, the grid will be discarded. If the sub-image has no CD8+ T cells in it, it will also be discarded.
- Most good patients have around 300 sub-images, some have more than 1000 sub-images, some just have 80 images. Poor patients have fewer sub-images.

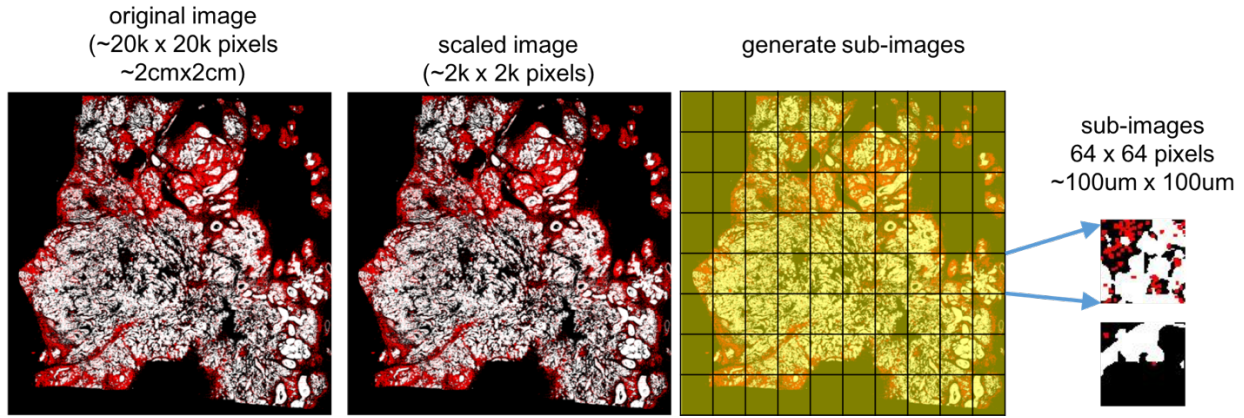
Figure 1: Illustration of the image pre-processing procedure.

Training
- The machine learning approach adapted in this work is Mxnet from: https://github.com/theislab/deepflow
- Each sub-image is the input of each training process. After each training process, we assign the output to be good or poor.
- For each patient, 80% of all its sub-images (~240) are used for the training process, whereas the remaining are used for prediction.
- There are more sub-images for patients with good outcome (4 times) than ones with poor outcome. We made copies of sub-images for patients with poor outcome to make the balance of the training samples.
- In total, we use about 11000 sub-images for the training process. The logics of the training process is as follows: there is a set of parameters in the Mxnet. With one input sub-image, a probability is given by Mxnet to say this sub-image is viewed as an image from a patient with good outcome. Since we know a prior that where this sub-image comes from, based on the difference between this probability and its known value (0 for good outcome or 1 for poor outcome), Mxnet can update its internal parameters automatically. 1 EPOCHS is defined as the process in which all sub-images were served as the input and the internal parameters of Mxnet were updated. We run 100 EPOCHS to train the Mxnet. If we select the cut-off probability to be 0.5, the converging average probability of Mxnet for ~11000 training sub-images is 0.85, which is smaller than 1. This is reasonable because good patients also have some parts which look like a poor patient, and vice versa.

| patient number (good outcome) | Number of sub-images | patient number (poor outcome) | Number of sub-images |
|---|---|---|---|
| 21 | 80 | 1 | 229 |
| 22 | 1011 | 2 | 128 |
| 23 | 1174 | 4 | 198 |
| 26 | 345 | 7 | 66 |

| 27 | 1153 | 8 | 177 |
|----|------|----|-----|
| 28 | 449 | 12 | 199 |
| 30 | 324 | 13 | 148 |
| 33 | 123 | 16 | 243 |
| 34 | 103 | 17 | 375 |
| 35 | 367 | | |
| 36 | 758 | | |
| 37 | 337 | | |
| 38 | 554 | | |
| 39 | 86 | | |
| 40 | 349 | | |

Table 1: A detailed table of the number of sub-images for each patient.

**Results**

- We use the output probability from mxnet to predict the test sub-image to be from patients with good or poor outcome. For each sub-image left for test, the threshold in the prediction is set as 0.5.
- Next, we calculate the fraction of sub-images that is predicted to be from patients with good outcome for each patient.
- Based on the fraction, we can select a cut-off value to predict a patient to have a good or poor outcome. The accuracy of our prediction is a function of this cut-off value, which is shown in here:

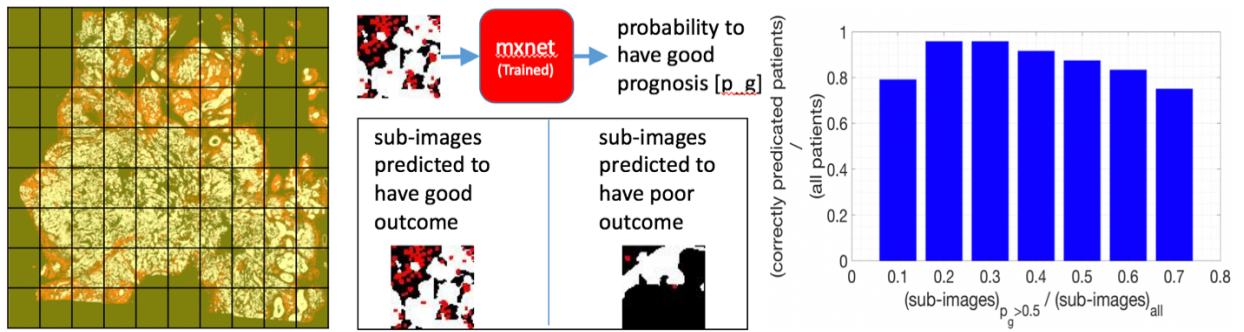

Figure 2: Fraction of the correctly-predicted patients as a function of the cut-off percentage of "good" sub-images in the prediction set.

| 0.1 | 19/24(number 2,4,7,13,17) |
|-----|---------------------------|
| 0.2 | 23/24(number 4) |
| 0.3 | 23/24(number 4) |
| 0.4 | 22/24(number 21 and 40) |

| 0.5 | 21/24(number 21,39,40) |
| --- | --- |
| 0.6 | 20/24(number 21,35,39,40) |
| 0.7 | 18/24(number 21,23,35,38,39,40) |

Table 2: A detailed table shows the specific patients that are wrongly predicted for a given cut-off of the fraction of "good" sub-images.

- We find 0.3(or 0.2) could be a good threshold for final decision. It means if there are more than 30 percent test sub-images are predicted to be from patients with good outcome, then the patient will be labeled as good, if the number is smaller than 30 percent, the patient will be labeled as poor. We apply this overall decision process on each patient and get 23/24 accuracy. Patient 4 will be predicted as a good patient, which is wrong.
- We tried to show what the machine learns about the pattern
    - The detailed characterization of the red-to-white ratio for sub-images in the training set.

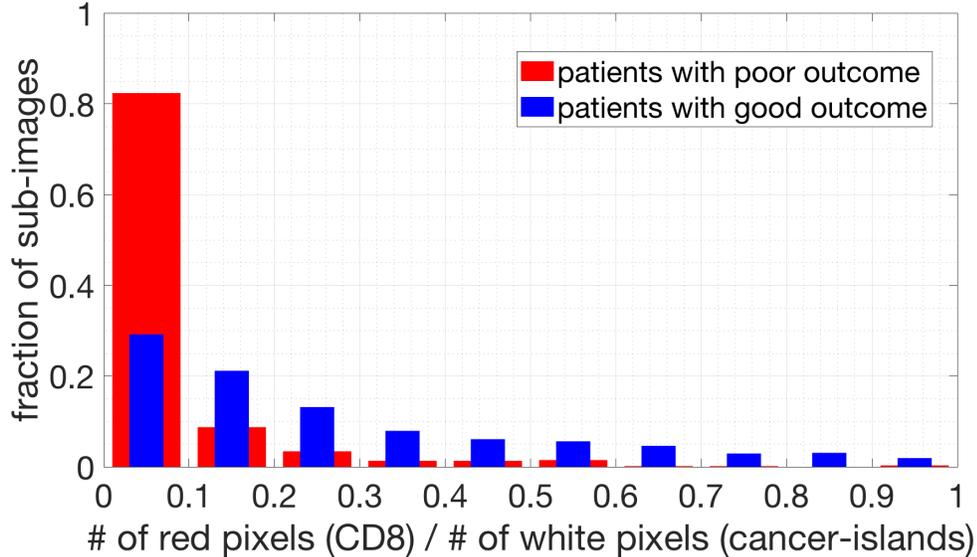

Figure 3: Sub-images from patients with good outcome have higher CD8+/cancer ratio. In this plot, the x-axis is the ratio between T cell pixels over cancer island pixels, y-axis is the fraction of sub-images. Results for patients with poor/good outcome is in red/blue, respectively.

- Sampled sub-images in the prediction subset to illustrate the relationship between predicted-probability and CD8-cancer pattern.

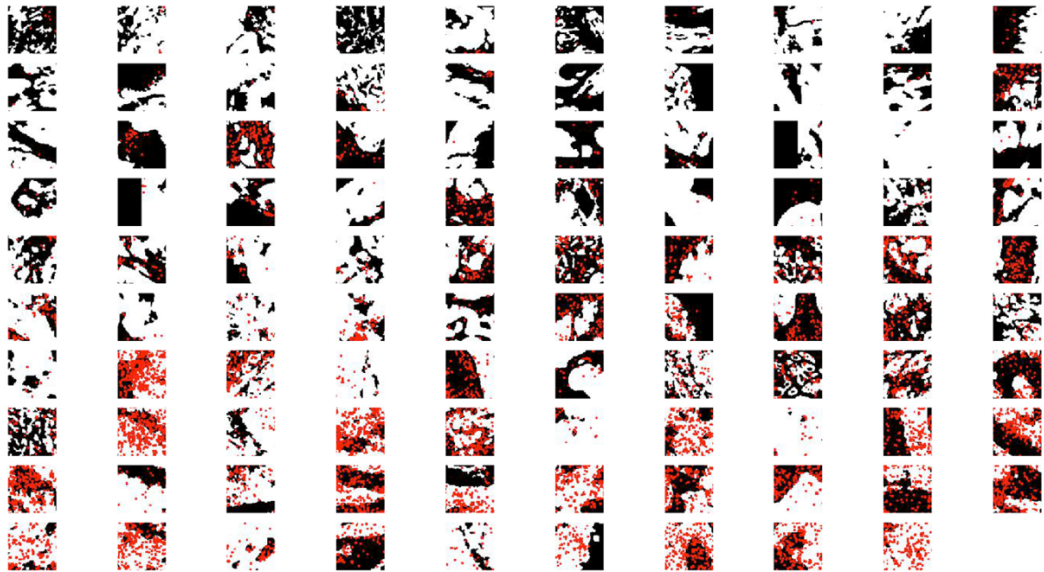

Figure 4: Sampled sub-images with increasing probability predicted to be "good". In this figure, from top to bottom, from left to right, the possibility predicted to be good is increasing.

[1] Heidari M, Khuzani A Z, Hollingsworth A B, et al. Prediction of breast cancer risk using a machine learning approach embedded with a locality preserving projection algorithm[J]. Physics in Medicine & Biology, 2018, 63(3): 035020
[2] Agarap A F M. On breast cancer detection: an application of machine learning algorithms on the wisconsin diagnostic dataset. https://arxiv.org/abs/1711.07831
[3] Saltz J, Gupta R, Hou L, et al. Spatial organization and molecular correlation of tumor-infiltrating lymphocytes using deep learning on pathology images[J]. Cell reports, 2018, 23(1): 181.
[4] Mahmoud S, Lee A, Ellis I, et al. CD8+ T lymphocytes infiltrating breast cancer: A promising new prognostic marker?[J]. Oncoimmunology, 2012, 1(3): 364-365.
[5] https://www.ncbi.nlm.nih.gov/pmc/articles/PMC5761898/pdf/pone.0190158.pdf